\documentclass[sigconf]{acmart}

\AtBeginDocument{%
  \providecommand\BibTeX{{%
    \normalfont B\kern-0.5em{\scshape i\kern-0.25em b}\kern-0.8em\TeX}}}

\copyrightyear{2024}
\acmYear{2024}
\setcopyright{acmlicensed}\acmConference[CHI '24]{Proceedings of the CHI Conference on Human Factors in Computing Systems}{May 11--16, 2024}{Honolulu, HI, USA}
\acmBooktitle{Proceedings of the CHI Conference on Human Factors in Computing Systems (CHI '24), May 11--16, 2024, Honolulu, HI, USA} \acmDOI{10.1145/3613904.3642248} \acmISBN{979-8-4007-0330-0/24/05}

\usepackage[utf8]{inputenc}
\RequirePackage[T1]{fontenc}
\RequirePackage[tt=false, type1=true]{libertine}
\RequirePackage[varqu]{zi4}
\RequirePackage[libertine]{newtxmath}

\title[Unpacking Sustainability in Physicalization Practices]{From Exploration to End of Life:\texorpdfstring{\\Unpacking Sustainability in Physicalization Practices}{}}
\usepackage{color}
\usepackage{soul}
\usepackage{tikz}

\newcommand{\mention}[2]{\textit{``#1'' \small{(\textbf{#2})}}}

\newcommand{\rev}[1]{\textcolor{black}{#1}}

\DeclareRobustCommand{\icon}[2][height=1.5]{%
  \begingroup\normalfont
\includegraphics[#1\fontcharht\font`\B]{#2.png}%
  \endgroup
}

\definecolor{intent-color}{HTML}{88109D}
\definecolor{impact-color}{HTML}{D7AC12}

\definecolor{planes-c}{HTML}{12CDD4}
\colorlet{planes-color}{planes-c!30}
\definecolor{badges-c}{HTML}{FAC710}
\colorlet{badges-color}{badges-c!30}
\definecolor{plants-c}{HTML}{0CA789}
\colorlet{plants-color}{plants-c!30}
\definecolor{pandemic-c}{HTML}{CEE741}
\colorlet{pandemic-color}{pandemic-c!30}
\definecolor{bicycle-c}{HTML}{DA0063}
\colorlet{bicycle-color}{bicycle-c!30}
\definecolor{water-c}{HTML}{808080}
\colorlet{water-color}{water-c!30}
\definecolor{physical-c}{HTML}{9510AC}
\colorlet{physical-color}{physical-c!30}
\definecolor{plastic-c}{HTML}{414BB2}
\colorlet{plastic-color}{plastic-c!30}

\newcommand{\iconcircle}[1][black, fill=black]{\tikz[baseline=-0.5ex]\draw[#1,radius=4.5pt] (0,0) circle ;}%

\newcommand{\project}[2][black]{\iconcircle[#1, fill=#1]\!\!\!\!\textit{#2}}

\newcommand{\dimtitle}[2]{\textbf{\textcolor{#1}{#2}}}

\begin{document}

\author{Luiz Morais}
\email{lamm@cin.ufpe.br}
\orcid{0000-0002-5506-9473}
\affiliation{
  \institution{Universidade Federal de Pernambuco}
  \city{Recife}
  \country{Brazil}
}

\author{Georgia Panagiotidou}
\email{georgia.panagiotidou@kcl.ac.uk}
\orcid{0000-0003-4408-6371}
\affiliation{
  \institution{King's College London}
  \city{London}
  \country{United Kingdom}
}

\author{Sarah Hayes}
\email{sarah.hayes@mtu.ie}
\orcid{0009-0004-1463-1361}
\affiliation{
    \institution{Munster Technological University}
  \city{Cork}
  \country{Ireland}
}

\author{Tatiana Losev}
\email{tatiana.losev@sfu.ca}
\orcid{0000-0002-0363-8072}
\affiliation{
    \institution{Simon Fraser University}
  \city{Surrey}
  \country{Canada}
}

\author{Rebecca Noonan}
\email{rebecca.noonan@mycit.ie}
\orcid{0009-0000-8139-6657}
\affiliation{
    \institution{Munster Technological University}
  \city{Cork}
  \country{Ireland}
}

\author{Uta Hinrichs}
\email{uhinrich@ed.ac.uk}
\orcid{0000-0001-7494-0941}
\affiliation{
    \institution{University of Edinburgh}
  \city{Edinburgh}
  \country{United Kingdom}
}

\renewcommand{\shortauthors}{Morais et al.}
\begin{abstract}
Data physicalizations have gained prominence across domains, but their environmental impact has been largely overlooked. This work addresses this gap by investigating the interplay between sustainability and physicalization practices. We conducted interviews with experts from diverse backgrounds, followed by a survey to gather insights into how they approach physicalization projects and reflect on sustainability. Our thematic analysis revealed sustainability considerations throughout the entire \textit{physicalization life cycle}---a framework that encompasses various stages in a physicalization's existence. Notably, we found no single agreed-upon definition for sustainable physicalizations, highlighting the complexity of integrating sustainability into physicalization practices. We outline sustainability challenges and strategies based on participants' experiences and propose the Sustainable Physicalization Practices (SuPPra) Matrix, providing a structured approach for designers to reflect on and enhance the environmental impact of their future physicalizations.
\end{abstract}



\date{May 2023}

\maketitle

\section{Introduction}

Data physicalizations have emerged as versatile tools with wide-ranging applications in diverse domains, including data analytics, communication, education, and accessibility \cite{dragicevic2020data}. The process of conceptualizing a data physicalization entails a complex interplay of decisions, encompassing the selection of appropriate materials and the consideration of techniques for constructing and assembling its components \cite{bae2022making}. Considering the environmental footprint and impact associated with physicalizations, the importance of discussing environmental sustainability in this field has become increasingly evident. While the CHI community has initiated discussions on potential directions towards sustainability in Data Physicalization \cite{sauve2023physicalization}, no comprehensive work has systematically explored how to promote sustainability and embed it into future practices, leaving an essential aspect of this field uncharted. This work contributes to filling this gap by exploring the role of sustainability within the different stages of physicalization practice.

To gain insights into sustainability in Data Physicalization, we conducted a comprehensive study with experts. We first interviewed artists, academics, and designers from different domains. Then, we leveraged the responses obtained from the interviews to design a survey, reaching out to a broader and diverse group of participants. Through thematic analysis \cite{braun2012thematic}, we explored participants' physicalization practices from the perspective of environmental sustainability. What defines a sustainable physicalization? What are the prevailing attitudes and obstacles regarding physicalization sustainability? What factors affect the process of creating a physicalization? Our findings highlight sustainability considerations across the entire \textit{physicalization life cycle}---which refers to the various stages throughout a physicalization's existence, spanning from the \textit{exploration} phase, where designers gather inspiration, to the \textit{end of life}, where decisions are made about handling, disposal, or reuse of remaining physical components. Our approach aligns with other studies exploring life cycles of garments~\cite{gwilt2020practical} and sustainable making practices \cite{bell2022reclaym}. 

We did not find a universally accepted definition for sustainable physicalization, nor did we discover an objective set of guidelines to ensure sustainability. Our findings indicate that incorporating sustainability into physicalization practice is a complex undertaking that requires critical reflection and proactive measures throughout the entire life cycle of the physicalization to reduce its environmental impact. This work highlights the challenges \rev{encountered by participants in integrating sustainability into} their projects and the strategies they employed \rev{or considered afterward} to promote sustainability within their practice. Drawing on our research findings and existing HCI literature, we have formulated ten sustainability dimensions for physicalization practice. These dimensions combined with a set of prompting questions form a call-to-action for promoting sustainability in future physicalization practices.

This paper introduces the research field of \textit{Sustainable Data Physicalization} (SDP), which lies at the intersection of Sustainability, Data Physicalization, and Sustainable HCI. SDP has emerged from the recent interest of the Data Physicalization community in making environmental sustainability a focus of research, practice and teaching. We consider SDP as a broad discipline that encompasses research focusing on using data physicalization to raise awareness about environmental issues (e.g., \cite{sauve2020econundrum,stegers2022ecorbis,sauve2023edo,lindrup2023carbon}) as well as research that advocates for sustainability within the physicalization practice (e.g., \cite{hayes2023zero,baur2023location,lindrup2023data, kanis2023engaging}). Similar to Sustainable HCI, this discipline seeks to understand and deal with the impact of technology on the environment. However, Sustainable Data Physicalization more specifically focuses on sustainability concerns that are particular to the process of transforming data into a physical representation.

This work contributes to Sustainable Data Physicalization in multiple ways:

\begin{itemize}
    \item \textbf{It introduces the physicalization life cycle}, which serves as a framework to outline the key phases in the lifespan of a physicalization. This framework can enable designers to consider sustainability throughout the stages of \textit{exploration} (\icon[height=1.1]{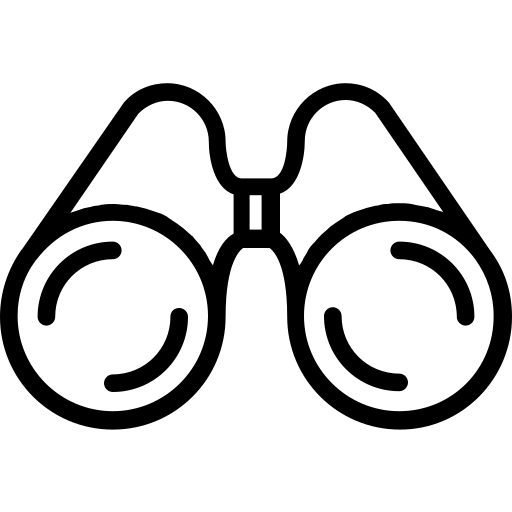}), \textit{ideation} (\icon[height=1.1]{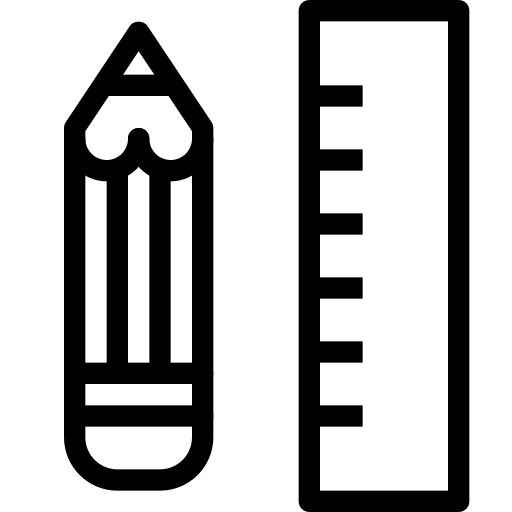}), \textit{creation} (\icon[height=1.1]{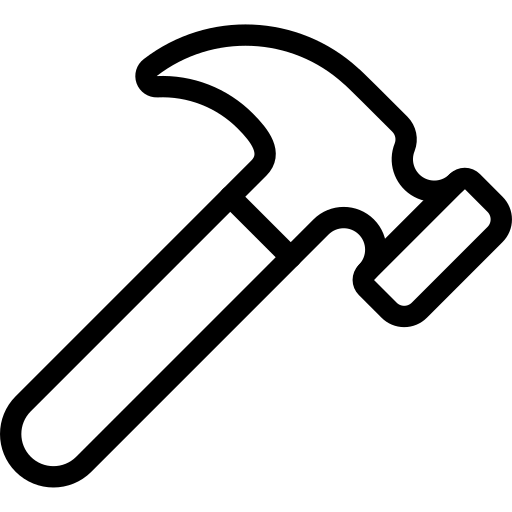}), \textit{presentation} (\icon[height=1.1]{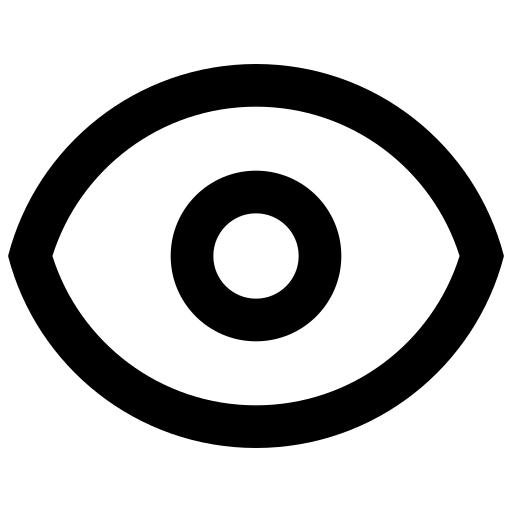}), and \textit{end of life} (\icon[height=1.1]{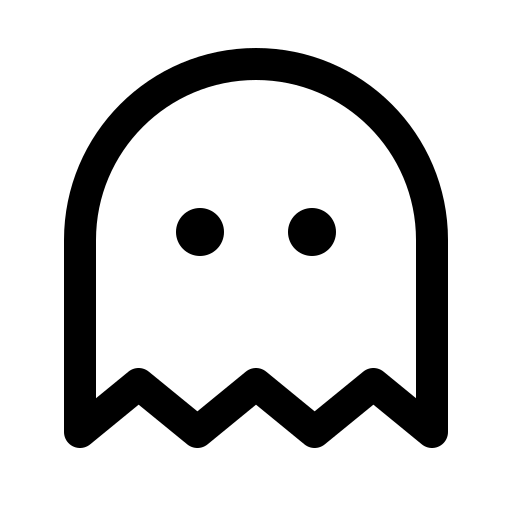}).
    \item \textbf{It outlines sustainability challenges and strategies} based on participants' experiences. Challenges encompass cost, material constraints, ownership, aesthetics, transportation, reuse, and longevity. Strategies involve early sustainability reflection, alternative materials/practices, transportation facilitation, engagement promotion, and recognizing physicalization's end-of-life.
    \item \textbf{It characterizes sustainability dimensions for physicalization practice}, drawn from our findings, experiences, and the literature. These dimensions encompass designers' intentions to anticipate, negotiate, inspect, and reflect on sustainability, as well as the impact of their design choices on materiality, longevity, versatility, consistency, visibility, and viability of physicalizations.
    \item Finally, \textbf{it issues a call-to-action} by \rev{proposing the Sustainable Physicalization Practices (SuPPra) Matrix, consisting of} prompting questions related to sustainability across different dimensions and life cycle phases of a physicalization. How long does a physicalization need to last? How can a physicalization be used by more people? Where and how can the materials be stored, disposed of, recycled or reused? These questions are a starting point to motivate designers to critically reflect on sustainability.
\end{itemize}

\rev{Our contributions serve to both open the conversation around sustainability and data physicalization, at a community and individual level, and to provide an initial structure and language for sustainability-related data physicalization concerns and activities.}
\section{Background}

While the idea of making data tangible is as old as human kind, recent advancements in digital fabrication, tangible interfaces, and shape-changing displays have fueled research in data physicalization~\cite{jansen2015opportunities}. Within this domain, academics and practitioners explore innovative approaches to transform abstract data into tangible and interactive artifacts. Recent work in the area has started to focus on exploring design processes~\cite{huron2022making} and the impact of data physicalization in terms of perception and meaning-making in different contexts (e.g., self-reflection~\cite{thudt2017data}, energy~\cite{hansen_lumen_2020}, and civic participation~\cite{claes_tangible}, just to name a few). However, \rev{the creation, control, modification, maintenance, and distribution of tangible artifacts can pose sustainability challenges for the environment if not properly addressed~\cite{holmquist2023bits}. Therefore, it is crucial to explore the environmental sustainability of current physicalization practices.} This section aims to provide an overview of the current state-of-the-art in data physicalization design, in light of current discussions of sustainability in HCI and Design.

\subsection{\rev{Sustainable HCI}}

The debates surrounding sustainability in Interaction Design and HCI emerged in the 2000s. Blevis~\cite{blevis2007sustainable} introduced the discussion about environmentally conscious practices in Interaction Design, emphasizing the importance of establishing a strong link between the creation and eventual disposal of artifacts as well as of advocating for the integration of renewal and reuse as essential elements within the design process. Mankoff et al.~\cite{mankoff2007environmental} further highlighted sustainability concerns within the CHI community, proposing two approaches: (1) sustainability \textit{through} design to increase awareness of environmental issues and foster sustainable behavior, and (2) sustainability \textit{in} design, promoting eco-friendly practices like waste reduction and device reuse. The term \textit{Sustainable HCI (SHCI)} was coined by DiSalvo et al. in 2010~\cite{disalvo2010mapping}, who outlined the research field, unveiling key strands including persuasive technologies for behavior change, formative studies to understand users' attitudes towards sustainability, and works that critically rethink HCI methods for promoting sustainable practices. Sustainable HCI has established itself as a research field, developing and embracing an evolving agenda towards environmental improvement based on the Sustainable Development Goals and beyond ~\cite{bremer2022have,hansson_decade_2021}. 

The SHCI community faced critique for its predominant focus on persuasive technologies~\cite{disalvo2010mapping,preist2013post}, but it has been shifting its research efforts towards more qualitative and speculative approaches~\cite{bremer2022have,asgeirsdottir_making_2023}. The early research on \rev{SHCI} mainly focused on approaches such as the evaluation of individuals' levels of sustainable behavior based on metrics defined by designers (e.g.,~\cite{bangpowerhouse}) or the eco-feedback about consumption, waste or other aspects related to sustainability (e.g.,~\cite{al2007iparrot,gustafsson2005power}). However, as Brynjarsdottir et al. argued, focus on persuasive technologies can overlook critical social dynamics inherent in addressing environmental issues, and design studies usually lack evidence for long-term environmental impact~\cite{brynjarsdottir2012sustainably}. Recent efforts in Sustainable HCI have started to focus on innovative design processes intended to reduce environmental impact, including the use of biodegradable materials, such as biofoam~\cite{lazaro2022exploring}, natural materials such as clay~\cite{bell2022reclaym} or leaves~\cite{song2022towards}, or even living organisms~\cite{bell2023scoby}. Additionally, the concepts of un-fabrication~\cite{wu2020unfabricate} and unmaking \cite{song2021unmaking} also shed light on the value of reusing or evolving existing artifacts. These Sustainable HCI endeavors hold the potential to promote responsible and environmentally conscious design practices.

\subsection{\rev{Data Physicalization Practices and Sustainability}}

Designers and researchers have been exploring the growing space of data physicalizations \cite{djavaherpour2021data,bae2022making}\rev{, including the elements that constitute its ecology~\cite{sauve2022physecology}}. They integrate principles from Data Visualization, Interaction Design, and Human-computer Interaction to create a myriad of artifacts \rev{that represent data through their physical properties}. The resulting artifacts can be built based on handcraft techniques like knitting or pottery \cite{thudt2017data} or more technology-oriented approaches such as laser-cutting \cite{gourlet2017cairn} or 3D printing \cite{barrass2016diagnosing}. Data physicalizations range from ephemeral data sculptures made out of ice \cite{segal2023grewingk} to long-lasting artifacts that can be part of a public space for years \cite{segal2023tidal}. There are also physical representations that dynamically actuate as data changes \cite{taylor2015data} or participatory physicalizations made out of beach debris \cite{klauss2023perpetual}. The materials used for creating data physicalizations also vary greatly depending on the project: from a mix of plants and electronic components used to create dataponics \cite{cercos2016coupling} to a kit with paper, elastics, and markers used to run the \textit{Let's Play with Data} workshop \cite{duarte2023let}. In summary, the design space of physicalizations is vast and includes different design practices.

The process of designing data physicalizations involves multiple stages, which can vary by project \cite{huron2022making}. Typically, papers on physicalizations outline steps like ideation, implementation, and deployment (e.g., \cite{perovich2020chemicals}). Ideation generates ideas, enabling innovative concepts. Implementation translates ideas into tangible artifacts through design and development. Deployment brings creations into the physical world for experience. However, the process of conceiving data physicalizations can include additional or alternative phases depending on the project's context and objectives. For instance, Waldschutz and Hornecker \cite{waldschutz2020importance} discuss the importance of initially curating the data before starting to develop design concepts to avoid common pitfalls. Morais, Andrade, and Sousa \cite{morais2022exploring} also mention a prototyping phase prior to the installation of the final version of their physicalization. Additionally, workshops on physicalizations \cite{huron2017let} emphasize stages such as material selection and reflections on the physicalization process. While prior research has covered stages in data physicalization creation, a significant gap exists in addressing their connection to sustainability and offering deeper insights into what happens when the project ends. Thoroughly exploring the entire physicalization life cycle is vital for understanding sustainability issues.

A lot of work in Data Physicalization addresses the issue of sustainability directly through the physicalization artifact itself, raising awareness, for example, of issues such as pollution~\cite{offenhuber2019dustmarks} or plastic waste~\cite{sauve2020econundrum,sauve2023edo,klauss2023perpetual}. In line with the existing HCI literature, the Data Physicalization community has also recently embarked on exploring design processes to enhance the sustainability of data-driven physical artifacts. Notably, the CHI 2023 workshop on \textit{Physicalization from Theory to Practice}~\cite{sauve2023physicalization} played a pioneering role by sparking insightful discussions on sustainability in Data Physicalization research. The workshop delved into various aspects, such as the creation of a zero-waste physicalization kit~\cite{hayes2023zero}, the investigation of sustainable material choices, and the formulation of strategies for upcycling, disposal, and recycling of physical artifacts ~\cite{baur2023location, lindrup2023data, kanis2023engaging}. However, while the \rev{SHCI} community has addressed certain related topics, there are specific aspects unique to data physicalizations that still require attention. The research and practice of Sustainable Data Physicalization remain unexplored \cite{hayes2023zero}. Therefore, there is a pressing need for comprehensive work that systematically explores the interplay between physicalization practices and sustainability.

\subsection{\rev{Sustainability Frameworks}}

\rev{Several frameworks have been proposed in related works to assess sustainability. One widely recognized framework is the Life Cycle Assessment (LCA)~\cite{curran2013life}, extensively used across various disciplines to compare the environmental impact of products. This framework segments a product's life cycle into five key phases: raw material extraction, manufacturing and processing, transportation, usage and retail, and waste disposal. Operating as a standardized method, the LCA employs diverse metrics to measure the environmental sustainability of a product, offering a scientific basis for industries and governments. The HCI literature has also proposed sustainability frameworks over the years to evaluate computing systems. Dillahunt, Mankoff, and Forlizzi~\cite{dillahunt2010proposed} introduced a prescriptive framework consisting of yes/no questions related to aspects like energy consumption, user behavior, and technology reuse. Similarly, Toyama~\cite{toyama2015preliminary} presented an initial taxonomy of value for sustainable computing, suggesting that sustainability should be considered as a function of impact, intention, and effort. Expanding upon Toyama's work, Lundstrom and Pargman~\cite{lundstrom2017developing} proposed the Sustainable Computing Evaluation Framework (SCEF), aiming to offer a more practical tool for assessing the sustainability of computing systems. While these frameworks assist practitioners and researchers in objectively evaluating sustainability within their projects, they often present one-size-fits-all strategies that constrain the understanding of sustainability to a mere score or index.}

\rev{Part of the SHCI community advocates for encouraging reflection on sustainability rather than fixating on rigid rules and predefined evaluation models~\cite{remy2017limits}. Remy et al.~\cite{remy2018evaluation} introduced an evaluation model comprising five integral components: goals, mechanisms, metrics, methods, and scope. This model takes a generative approach, designed to empower individuals to devise personalized methods for evaluating technologies. By proposing this model, the authors encouraged the SHCI community to develop more nuanced and concrete approaches for evaluating sustainability, emphasizing context-specific methods over universal approaches. Grimal et al.~\cite{grimal2021design} took a similar approach and extended the SCEF framework to consider sustainability not as a goal but as a learning process. The authors proposed adaptations to the SCEF framework to motivate engineers reflect on sustainability issues within computer-based projects. The changes involve considering the impact of systems throughout the entire product life cycle and the temporality and multipliticy of stakeholders in the process. The authors argue that ``this updated model is relevant both for the evaluation of disruptive projects to know their `level' of sustainability, but also to accompany project leaders (stakeholders in general) for a change in their way of thinking and conceiving the technology.''. Likewise, Lazaro, Wang, and Vega~\cite{lazaro2020introducing} modified the Life Cycle Assessment (LCA) framework to suit the unique context of digital fabrication, presenting the Sustainable Prototyping Life Cycle for Digital Fabrication. This framework targets decision makers, intending to raise their awareness about sustainability impacts in prototyping stages, considering alternative materials and eco-friendly practices. Finally, the Art community \cite{baadger2020guide} has also introduced a guideline featuring prompting questions designed to assist artists in reflecting on their creative processes. While these frameworks and guidelines offer insight into sustainability within various domains, they often overlook specific aspects crucial to Data Physicalization. Hence, there is a pressing need for a specialized framework addressing the life cycle of physicalizations.}
\section{Study Design: Semi-Structured Interviews and Online Survey}

\rev{Building upon the emerging exploration of sustainability within HCI, our objective is to provide a comprehensive perspective on sustainability in Data Physicalization. To achieve this goal, we} explored the dynamic relationship between sustainability and the practices of physicalization \rev{experts}, as perceived through their own experiences. By recognizing that designing physicalizations is often iterative, we framed this design process as a \textit{life cycle}, mirroring the cyclical and material nature of physicalizations themselves. To investigate this relationship, we initially conducted interviews with \rev{a group of artists, academics, and designers from different domains}. Subsequently, we expanded our research scope by developing an online survey with similar questions. This survey allowed us to explore a wider array of contexts and gather insights from a broader range of \rev{participants}. In doing so, we sought to understand \rev{participants}' considerations, motivations, tensions, and opportunities within their individual practices. The research has been approved by our university's ethics committee.

\subsection{Instruments and Procedure}

We crafted our interview and survey questions in alignment with the phases of a physicalization's life cycle, with a particular focus on their relevance to sustainability concerns. Overall, the questions were related to considerations behind participants' \textit{design} choices, how the physicalization \textit{use} could impact the environment, and what were the plans for the physicalization once it reached its \textit{end of life}. We approached the topic of sustainability with sensitivity, refraining from making any assumptions about its relevance to the participants' design choices. Initially, we introduced sustainability-related topics, such as disposal, reuse, and material selection, without explicitly discussing the definition of sustainable physicalization practices. Subsequently, we posed more direct and explicit questions about sustainability. For more detailed information, the interview script and survey can be found in the supplementary materials.

\subsubsection{Interviews}

Participants engaged in 45-60 minute interviews conducted by one of three researchers through online video calls. During the interviews, participants were asked about their definitions of \textit{data physicalization} to gain an understanding of how each participant perceives the term. To provide context, participants were encouraged to select one of their data physicalization projects they wished to discuss (\rev{the selected projects are shown in} Figure~\ref{fig:projects}), offering details on its purpose, intended audience, materials used, and its journey from inception to the present. To mitigate potential bias towards design aspects, we allowed participants to share any deviations in the life cycle of their data physicalization from our proposed interview framework. All interviews were recorded with participants' consent, transcribed, and subsequently utilized for thematic analysis and critical discussions by the research team.

\subsubsection{Survey}

After conducting the interviews, we proceeded to develop an online survey focused on the life cycles of data physicalizations. We used the interview script and responses as a foundation for elaborating the survey questions. We hoped the survey would capture a larger net of people to enrich the scope and characterization of the physicalization life cycles. To achieve this, we extended an open invitation to our survey across online communities, including research groups and email lists encompassing individuals with expertise in data visualization/physicalization and HCI. Participants were provided with the option to maintain anonymity or to share their contact information for potential follow-up inquiries. The survey remained open for responses for a period exceeding one month.

\subsection{Participants}

\rev{To recruit participants, we compiled a list of potential interviewees leveraging our network of colleagues and people interested in data physicalization from the DataPhys website\footnote{http://dataphys.org/wiki/People}}. We contacted interview participants who have documented work on data physicalization, while making a conscious attempt to vary them among dimensions such as physicalization type, discipline and (assumed) gender. The final participant pool (N=8) provides multiple contexts for our study including expertise in 3D printing, crafting, electronics, large-scale artifacts, ephemeral exhibitions, waste-based materials, and physicalization activities. We name participants from I1 to I8. In addition to the interviews, we received 12 responses to the survey from participants (N=12) who have used data physicalization in their work\rev{, including physicalization teaching, physicalizations of personal and public data, and physicalization toolkits}. We name them from S1 to S12. Table \ref{tab:participants} summarizes all interview and survey participants.

\begin{table}
\centering
\begin{tabular}{c|c|l}
    \hline
    ID & Gender & Domain \\
    \hline
    I1 & Female & Research \\
    I2 & - & Research\\
    I3 & Male & Research \\
    I4 & Female & Industry, Research\\
    I5 & Female & Design, Research \\
    I6 & Female & Art \\
    I7 & Undefined & Design, Research \\
    I8 & Female & Art, Science \\
    S1 & Female & Research \\
    S2 & Male & Research\\
    S3 & Male & Research \\
    S4 & Female & Design, Research\\
    S5 & Female & Industry, Art \\
    S6 & Male & Art\\
    S7 & Male & Industry\\
    S8 & Male & Design\\
    S9 & Female & Research\\
    S10 & Female & Research\\
    S11 & Male & Research\\
    S12 & Male & Research\\
\end{tabular}
\caption{Interview and survey participant breakdown. Gender is self-defined.}
\label{tab:participants}
\end{table}

\subsection{Projects}

\begin{figure}[!ht]
    \centering
    \includegraphics[width=\columnwidth]{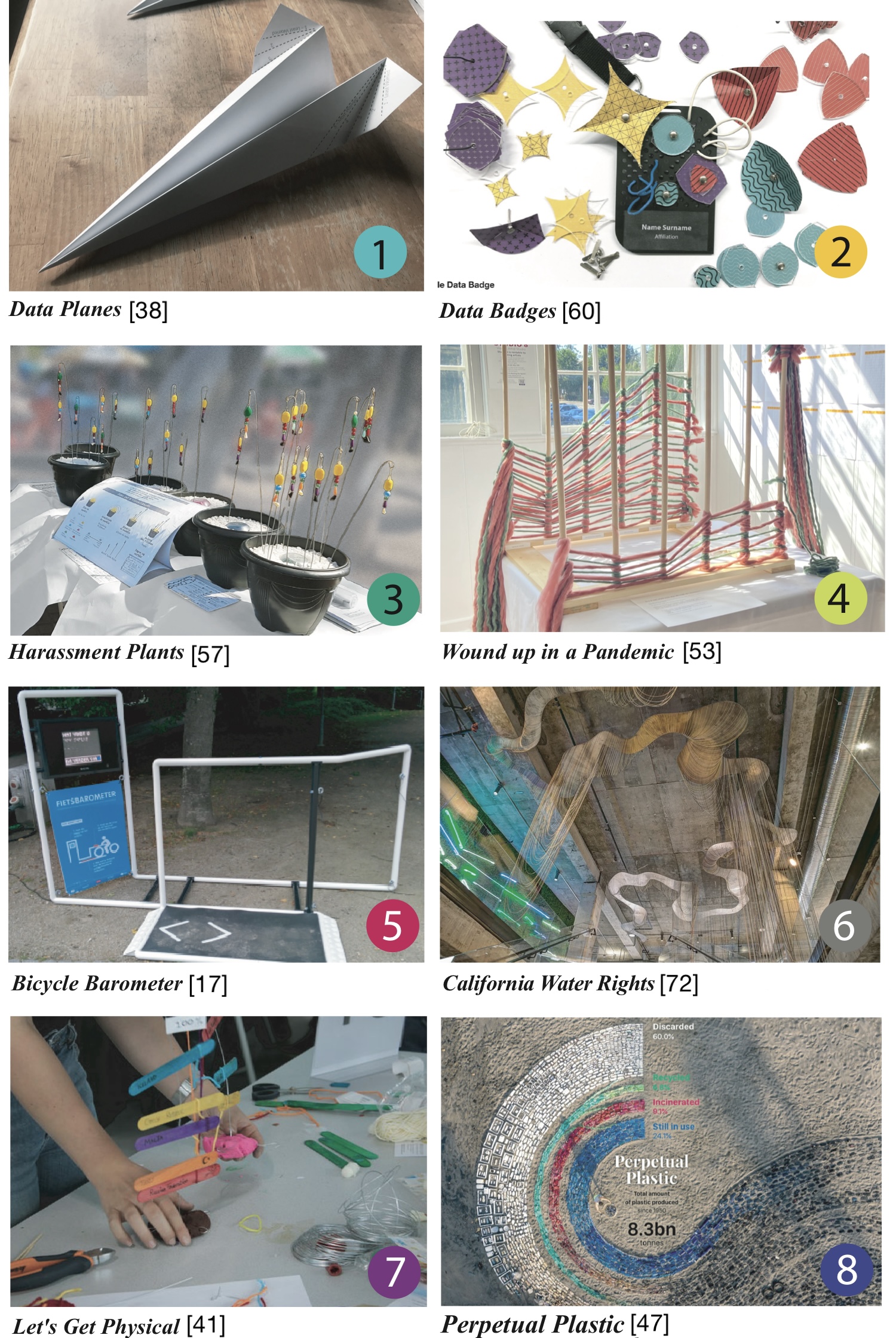}
    \caption{Projects from the interviews with physicalization experts. Credits: (1) Sarah Hayes; (2) Giorgia Panagiotidou; (3) Luiz Morais; (4) Tatiana Losev; (5) Sandy Claes; (6) Adien Segal / Photo: Mario Gallucci; (7) Samuel Huron; (8) Skye Moret, Liina Klauss, and Moritz Stefaner.}
    \label{fig:projects}
\end{figure}

\rev{Prior to the interviews, participants were prompted to select one of their projects that most aligns with this study's focus. All participants provided consent for descriptions of their projects to be included in academic publications, acknowledging that these descriptions might contain information that could potentially identify them as participants.} Our interview responses draw insights from eight distinct physicalization projects, \rev{ranging from artistic presentations, research studies, or workshops,} as depicted in Figure~\ref{fig:projects}. In the forthcoming findings and discussion sections, we will refer to these projects using numerical identifiers or their corresponding colors.

\begin{enumerate}
    \item \project[planes-color]{Data Planes} \cite{hayessarahthesis} is an exploratory, novel, paper-folding physicalization project, creating paper airplanes to represent data. Studies were deployed with Data Planes with parents and children working together to engage with data through this modality. Findings from these studies illustrate the potential for low-cost, accessible materials to add novelty and playfulness when engaging with data for both adults and children.
    \item \project[badges-color]{Data Badges} \cite{databadges2019} are do-it-yourself wearable physical representations of an individual's academic achievements and profile that are assembled using physical tokens attached to a wearable canvas. These badges are a creative and tangible way to showcase academic data, such as publications, citations, and collaborations, encouraging connections within academic communities.
    \item \project[plants-color]{Harassment Plants} represent sexual harassment stories in a public space and was conceived for a study \cite{morais2022exploring}. Each plant represents a specific incident and is made from cheap materials found in local stores. The stem's height indicates the approximate time of occurrence, while beads on the plants represent different aspects of the stories.
    \item \project[pandemic-color]{Wound up in a Pandemic} \cite{woundup} is an interactive data-based installation. The physicalization is constructed from personal data from participants answering a question related to the trust they held in sources of information during the COVID-19 pandemic by looping coloured string around wooden posts.
    \item \project[bicycle-color]{Bicycle Barometer} \cite{claesbicycle2016} is an interactive, public data display directed at urban cyclists. The display's goal is to enhance the cycling experience by presenting real-time cycling behavior data to embolden engagement among the cycling community.  Questions are presented on a LED display, encouraging passing cyclists to vote by interacting with the floormat, selecting a smiley face.
    \item \project[water-color]{California Water Rights} \cite{californiawaterrights2023} is a large-scale data sculpture that depicts the usage rates of water allocated to California. To encode the data, 1,071 strands of ball chain were used, with each ball representing approximately 326,000 gallons of water used in the state. The aim is to elicit opinions related to society's use of the resource of water.
    \item \project[physical-color]{Let's Get Physical} \cite{huron2017let} is a three-card, instructional, physicalization toolkit used within a workshop environment to counteract pitfalls the authors had been encountering in previous workshops. The researchers found that by amalgamating the cards into the workshop, they were able to avoid the previously observed issues. Participants were able to start the design phase more efficiently as the generation and preparation of data was provided as well as a structure to follow throughout.
    \item \project[plastic-color]Perpetual Plastic \cite{klauss2023perpetual} was displayed on a beach in Bali and consisted of plastic beach waste to indicate the quantity and consequence of marine debris. The physical items used in the display were not only representative of the data, it was the data itself collected from beach cleanups with volunteers in the locality. A total of 4,760 pieces of debris were included in the installation.
\end{enumerate}

\subsection{Data Analysis}

Three authors conducted thematic analysis on participant responses, using a combination of asynchronous collaboration on a digital whiteboard and online video-calling co-working sessions. Initially, they collectively created a codebook using a deductive approach, drawing from their expertise in physicalization practices, research, and interview questions. The initial codebook covered codes like project stakeholders, phases in the physicalization life cycle, materials, methods, outcomes, definitions, and sustainability-related aspects. In the second phase, the coders employed an inductive approach to develop new codes and themes guided by the interview data, engaging in discussions with the research group. Survey data were also analyzed inductively by two coders, aligning with existing interview-derived themes and identifying emerging topics, such as the `Waste and Resource Management' sub-theme related to digital prototyping (see Section \ref{subsubsec:embrace_alternative}).

\subsection{Reflexivity}

\rev{Our team comprises Data Physicalization researchers with diverse academic backgrounds (i.e., PhD students, postdocs, and professors) and sustainability expertise, all having engaged in physicalization projects as designers or consultants. Spanning across North America, South America, and Europe, our geographic diversity allows us a multifaceted lens through which to contemplate sustainability, considering the realities of different countries and cultures. This shared experience forms the basis for comprehensive discussions on the various phases, challenges, and opportunities within the physicalization life cycle. Through extensive critical discussions, drawing from our roles as both researchers and designers, we aim to deepen our understanding of sustainability and the interview data. Our forthcoming findings will address our interpretations and limitations while advocating for wider participation of practitioners in Sustainable Data Physicalization, fostering knowledge sharing and collaboration in this field.}

\section{Findings}

Our findings are based on themes that emerged from the interviews and survey responses, which shed light on sustainability in data physicalization practices.

\subsection{The Physicalization Life Cycle}

We began our interviews using a physicalization \textit{life cycle} framework, initially dividing it into \textit{design}, \textit{usage} and \textit{end of life} stages, drawing from our own experiences and existing literature \cite{huron2022making}. However, the insights from our participants prompted us to refine this model. They provided more detailed breakdowns, splitting \textit{design} into \textit{exploration}, \textit{ideation} and \textit{creation}. We also changed the name of the \textit{usage} phase to \textit{presentation} to better reflect the interviewee's experiences. These phases aren't always linear and can vary depending on project goals, but they all offer opportunities to incorporate sustainable practices. Our revised physicalization life cycle is outlined as follows:

\begin{itemize}
    \item[\icon{exploration}] In the \textbf{exploration}, a designer gathers information about the topic, the space, and other sources that may help them to get inspiration for the physicalization design. I6, for example, describes their routine during the exploration: \mention{So, usually my work actually starts with a really specific personal experience in a landscape. That leads to a bunch of research to identify and learn more about how natural phenomena forces are at play in that landscape. And then figuring out how to identify which dataset best tells that story or gets that concept across.}{I6}.

    \item[\icon{ideation}] In the \textbf{ideation}, a designer creates sketches, chooses materials, and establishes design constraints. I2 describes their practice: \mention{My design was very iterative, so [...] I started out sketching ideas with colors and things that could be a cool way to represent personal data. So even though I was sketching, it was meant to be three-dimensional already. I didn't have really any constraints, except that I wanted it to be something people could wear.}{I2}

    \item[\icon{creation}] In the \textbf{creation}, a designer creates prototypes, tests materials and the overall structure, and constructs the final physicalization. I1 explains how this phase is important to test the concept: \mention{There was a little bit prototyping [...] They've made three or four different ideas for how the planes would work. I'm kind of testing them out because I was also interested in how their flight patterns could be used to encode data}{I1}. This phase helps \mention{figuring out what materials and like all that to make the finished piece. And then obviously the fabrication of the piece as a whole}{I6}.

    \item[\icon{presentation}] In the \textbf{presentation}, the physicalization is transported, installed, and documented. I6 explains how this phase can happen: \mention{The installation is in terms of a site-specific project or a show. When you talk about data visualization, there's always the presentation aspect like how does it get out to the world? So that often comes in the form of a video documentation, a photographic documentation, or a presentation to the public through lectures. Or, in the case of artwork, it's actually shown in a space. In the case of public art, it's permanently installed somewhere. So those all are different threads of what happens after it's made.}{I6}

    \item[\icon{end-of-life}] Lastly, the \textbf{end of life} involves decisions regarding the disposal of, reuse, or storage of the physicalization. People sometimes decide to reuse their physicalizations like I7, who \mention{still us[es] [the physicalizations] in [their] class}{I7} or I3, who \mention{used the pots [from their physicalization] for a while to plant stuff.}{I3}. Others, like I5, store the physicalization: \mention{it's still in the basement available for usage}{I5}. Alternatively, some physicalizations are discarded after presentation: \mention{It was much faster to take it down. It took maybe 2 hours and the beach was [left] just how it was before.}{I8}

\end{itemize}

Our proposed physicalization life cycle offers a framework for considering sustainability in Data Physicalization. This model enables designers to reflect on sustainability principles by delineating key phases. In the following sections, we highlight the reflections that emerged throughout our participants' physicalization life cycles.

\subsection{What is a Sustainable Physicalization?}

The idea of a \textit{sustainable physicalization} sounded strange to some participants. I6, for instance, mentions: \mention{Sustainable physicalization? [long pause] I don't know. I guess I don't normally put those two words together because [..] they don't seem innately related, unless your outset is to do something sustainable}{I6}. On the other hand, I3 suggests that there are various ways of considering the sustainability of a physicalization.

\begin{quote}       
    \mention{We could have a sustainable physicalization, for example, because it doesn't use any resources and it produces minimal waste. Or if it uses, for example, reusable materials and everything. So, it could be like an ephemeral physicalization. That could just come back to the environment without any harm. Or, on the other hand, it could be like an enduring physicalization that we could reuse it for multiple installations. So, I think there's not a single definition of sustainable physicalization.}{I3}
\end{quote}

Besides its definition, there seems to be a limited understanding of how to address sustainability in physicalization practices. I7 points out that \mention{I haven't found a good framework and a good way to understand it.}{I7}. Nevertheless, our findings suggest two overall factors may help researchers and practitioners reflect on physicalization sustainability: 

\subsubsection{\dimtitle{intent-color}{Intent}}\label{subsubsec:intent}

It seems that critically reflecting on and acting towards sustainability throughout a physicalization's life cycle is a central aspect that contributes to the sustainability of physicalizations. According to I3, \mention{we can only have a sustainable physicalization when we have intent}{I3}. We can foster sustainability, for example by \mention{consider[ing] it[s] [sustainability aspects] \rev{in the very} beginning.}{I2}. Negotiating the decisions with colleagues or clients is another way of promoting a sustainable life cycle. I5 gives us an example: \mention{I had this person who was working on a data story on waste. And she was using new materials to present that, so we pushed her to use waste materials.}{I5}. Reflecting on a physicalization's sustainability during its development is also crucial: \mention{it's important to think about making sure whatever you know the materials and the methods you're using to make [sure] that work isn't damaging [the environment]}{I6}. Finally, accounting for our practices afterwards also helps producing more sustainable physicalizations in the future: \mention{now I'm starting to think about the disposal or reuse or everything, but it's something that I'm incorporating into my design process}{I3}.

\subsubsection{\dimtitle{impact-color}{Impact}}\label{subsubsec:impact}

The degree to which the design choices impact the environment also seems to play a role in sustainability. For example, the material choice can contribute to sustainability since, according to a participant, \mention{a sustainable physicalization is made of renewable materials}{I4}. However, we must reflect beyond materiality since \mention{simply stating whether one uses recyclable and compostable materials is ultimately a limited view}{S11}. Another barrier to sustainability can be the audience reach. For instance, a physicalization created by I8 during the Covid-19 pandemic did not reach its intended audience due to event cancellations and reduced publication opportunities, resulting in material and resource wastage. Similarly, the short lifespan and limited versatility of some physicalizations might also affect sustainability. I5 noted that physicalizations made by their students are quickly discarded or \mention{disappear somewhere into an attic}{I5}. Besides that, the physicalizations can be \mention{so fragile and need a lot of knowledge to understand how to be set up}{I5}, which makes it hard to reuse them. Other factors such as the consistency of the physicalization with the data and its viability may also affect sustainability, as will be discussed in the next sections. 

In summary, our findings indicate that there is not a universally accepted definition of a \textit{sustainable physicalization}, nor is there an objective set of guidelines to guarantee sustainability. Instead, this research reveals the intricate nature of sustainability within Data Physicalization, emphasizing considerations at different stages of a physicalization's life cycle. In the next sections, we delve into the considerations our participants had regarding challenges (Section \ref{subsec:challenges}) and potential strategies (Section \ref{subsec:strategies}) that could promote sustainability.

\subsection{Sustainability Challenges in Physicalization Practice}
\label{subsec:challenges}

Regardless of the fact that eight (out of 12) participants from the survey considered the environmental impact of their physicalization, there are some challenges in promoting sustainability. These challenges stem from issues like cost, material constraints, ownership, aesthetics, transportation, reuse, and longevity of physicalizations. This section describes some practical challenges of SDP and explores how our participants attempted to strike a balance between making environmentally friendly choices and addressing real-world constraints.

\subsubsection{Paying the cost for sustainability}
\label{subsubsec:cost_sustainability}

Adopting sustainable practices is not easy, and not everyone can afford to do so. The limited \textbf{availability of sustainable materials}, for example, is an obstacle to promoting sustainability in data physicalization practices. According to I8, \mention{trying to find materials that are actually sustainable and can degrade or can be reused does take extra time, energy and effort}{I8}. The participant also gives us an example from one their projects: \mention{Making sure the dyes you're using are natural, not synthetic. They're hard to find, natural dyes}{I8}. \textbf{Monetary reasons} may also play a role in material choice. Sometimes, designers need to opt for more affordable solutions, even if those materials clash with their personal environmental values. I7, for example, explains why they chose plastic instead of wood: \mention{In the case of [a physicalization kit], it's made with plastic tokens. Uh, ideally [it should be made of] wood tokens, but I'm not rich enough for this one, and it was easier to produce the plastic ones than the wood ones.}{I7}. Similarly, S1 also discusses why cost affected their choice of a `less sustainable' material: \mention{I used acrylic yarn; I could have chosen natural fibers that would have been more sustainable to produce (but also more expensive to obtain)}{S1}. Hence, there's a need to discuss solutions that can tackle the challenges tied to both obtaining and affording sustainable materials.

\subsubsection{Dealing with material constraints and ownership}
\label{subsubsec:material_ownership}

While some designers are keen on integrating sustainability into their projects, sometimes they encounter external limitations that can discourage them from aligning their design choices with eco-friendly principles. The requirement to conform to existing \textbf{safety or structural regulations}, for example, means that some designers are often confined in terms of what materials they can select when designing for public spaces. I6, who often creates large-scale public displays as part of their practice, described the specific difficulties that working within this context bring with it in terms of working in a sustainable way: 

\begin{quote}
    \mention{When you're talking about public art scale work, it's much easier to achieve that in studio work, where I'm making a small sculpture that gets given to someone to put in their house. It's a different conversation than if you're working with structural engineers who need to have approved materials with engineered testing to make sure they will stand up over time because cities won't approve those permits unless they know what the material is.}{I6} 
\end{quote}

Designers can also face tensions regarding the \textbf{material choice} and \textbf{ownership} of their projects when they receive commissions from companies. I5, for instance, described how designing for a company changed the expectations of how ‘finished’ the artifact should look. While the designer intended to create a final version of their physicalization using the same components of the low-fidelity prototype, the company wanted a more polished version made of aluminum. The participant expressed their discontent because the negotiations did not end as expected: \mention{I think that's also because a company works in this way. It's like all or nothing, but I think that's not sustainable. That there could have been a step in between. That's something I regret often.}{I5}. Tensions regarding ownership may also happen in the physicalization's presentation. I8 explained that they intended to post one \rev{of} their physicalizations on social media to increase visibility but they were prohibited by the sponsor since they \mention{couldn't share any of the visualizations beforehand}{I8}. Similarly, designers can face sustainability issues related to ownership when the project reaches its end, since companies own the artifact and can follow unsustainable practices of disposal, for example. I6 speculates: \mention{They own [the physicalization] once it's completed and installed. So they could, in theory, the very next day be like `actually we don't like it' and take it down to throw it away.}{I6}. The same might happen when a physicalization workshop ends: \mention{if people did not like what they created with the [physicalization] kit, they might have thrown it away after the logging stage.}{S6}. Designers must take a proactive approach while collaborating with stakeholders to ensure sustainable practices throughout the physicalization life cycle.

\subsubsection{Finding a common ground between data encoding, aesthetics and sustainability}
\label{subsubsec:aesthetics_sustainability}

Balancing data encoding, aesthetics and sustainability presents a significant challenge while designing physicalizations. The need to create physical representations that match the data and are visually captivating often conflicts with the need to reduce environmental impact. I8 explained how they \mention{got into a little bit of an \textbf{aesthetic versus data integrity} battle a couple of times}{I8} since their colleague was more focused on the artifact's \mention{aesthetic sensory experience}{I8} and I8 wanted to \mention{retain data integrity}{I8}. In fact, the urge to create aesthetically-pleasing physicalizations may sometimes deter designers from reflecting on the impact of their choices on sustainability. I5 argues:
\begin{quote}
    \mention{So, especially in design, we have this tendency [...] to have polished-looking artifacts that use nice materials. But often [...] the concept is not good. And when that's not good, then it's such a waste to use expensive materials to make something that in its essence is not good yet.}{I5}
\end{quote}

As presented in Section \ref{subsubsec:material_ownership}, \textbf{aesthetic viewpoints} also cause conflicts related to material selection while dealing with stakeholders. Sometimes, partners prefer certain materials (like aluminium as mentioned before) that make the physicalization `look more polished' instead of using sustainable materials or reusing materials from the prototypes. Therefore, it is necessary to negotiate and find a balance between the aesthetic preferences and sustainability concerns.

\subsubsection{Managing the transportation of physical artifacts}
\label{subsubsec:coping_transportation}

The challenge of managing the transportation of physicalizations is quite complex. First of all, some physicalizations need to be created and presented in different parts of the world from where their creators live, which contributes to an increased \textbf{carbon footprint}, as I4 mentions: \mention{I would say geographical distance is an obstacle and having [...] to fly or drive.}{I4}. Similarly, when \textbf{large and heavy physicalizations} need to be transported in order to be presented, it requires a substantial amount of resources and effort, which also contributes to the carbon footprint and raises sustainability concerns. I3 admits that transporting one of their physicalizations to a public space \mention{wasn't very sustainable}{I3} because \mention{It was very heavy. [...] So I had multiple travels from the lab to the public space, so we had to transport it in different vehicles.}{I3}. The participant speculated that they could have rented a pickup to decrease the effort and environmental impact but they faced a financial problem, which relates to our discussions in Section \ref{subsubsec:cost_sustainability}. Moreover, the \textbf{fragility of physicalizations} is also a problem. I7 echoed this sentiment when said that \mention{the problem is, if you try to carry it around, you will destroy it}{I7}. The participant explains the problem of transporting components: \mention{It's a mess to move with them. [...] I remember the first time I had to do a full workshop with a lot of participants... I had to buy a special case, a suitcase for that, and I broke the suitcase on the way back.}{I7}. This fragility impacts the sustainability of physicalizations in that it poses a risk to the overall lifespan of the design, while also limiting its ability to be transported from place to place so that it can be viewed by a wider audience. Hence, designers need to plan in advance whether they have to transport the physicalization and how to do it as sustainably as possible.

\subsubsection{Dealing with the transience of data}
\label{subsubsec:transience_reuse}

The characteristic of encoding data is what differentiates physicalizations from other physical artifacts. However, materializing data can bring a series of challenges to sustainability due to their temporality and volatility. For example, I3 discussed sustainability issues in the design of a physicalization since the \textbf{dataset} they used for building the physicalization \textbf{changed over time}, causing the waste of previous prototypes and materials: \mention{The materials changed because of the data. Basically, as the data changed, we had to change the concepts. And the next concept didn't fit into the previous materials.}{I3}. The way that many physicalizations are built around data, \mention{resulting in one-off solutions tailored to a specific dataset, problem, or interaction}{S11}, can also affect sustainability since it \mention{limits whether the proposed technique can be reused in other applications/datasets/environments}{S11}. In I5's experience, the fragility of physicalizations, combined with the potential for their \textbf{data to become outdated}, contributes to the difficulty of reusing them beyond their initial purpose. While other physical artifacts can be presented throughout time without major concerns, physicalizations are \mention{often more fragile}{I5} and, since they typically represent a specific dataset, they \mention{should be updated}{I5}. Finally, the participant comments on the difficulty of maintaining physicalizations compared to other objects: \mention{I mean it's just much more temporary aspects to get it out of the cabinet and present it.}{I5}. Designers need to plan how to create physicalizations that can be easily updated or that can be useful regardless of their original dataset.

\subsubsection{Acknowledging the longevity of physicalizations}
\label{subsubsec:acknowledging_longevity}

Physicalizations are materializations of data and, as such, they tend to last for years. Therefore, it is paramount to consider the challenges associated with the longevity of physicalizations. A common issue related to this subject is \textbf{creating a physicalization without anticipating its end of life}. For example, I3 built a physicalization for a research study and did not plan what to do with the artifact afterward: \mention{I didn't have long-term plans for the physicalization. I didn't plan anything. I don't know. Maybe I was just thinking that I could store it somewhere. But yeah, I didn't have any plans of disposing of it or whatever.}{I3}. The participant explained that part of the physicalization was reused, but other components could not be reused and had to be discarded. Other designers acknowledged the longevity of physicalizations and even refused to create \textbf{long-lasting artifacts} because of ethical reasons: \mention{If the requirement is to last fifty years, I either don't do that project, or I get rejected by proposing something that's not going to last that long, or I have to figure out the best way to do that.}{I6}. Conversely, for designers like I1, long-lasting physicalizations can also be considered sustainable, depending on the context. The participant argues that designers should ask themselves about the physicalization's end goal while reflecting on the \textbf{tension between longevity and sustainability}: \mention{Is this an artifact that I want to exist for a really long time or is this something that I know is gonna have a relatively short life cycle or a relatively short existence? I only need it to last the hour or only need it to last the day for the workshop I'm running?}{I1}. The relationship between physicalization longevity and sustainability still seems to be messy and should be further explored.

\subsection{Strategies to Promote Sustainability in Physicalization Practice}
\label{subsec:strategies}

This section outlines strategies adopted by our participants with the intent of fostering sustainability throughout their physicalization life cycle. It serves as a starting point for researchers and practitioners, offering insights to consider and build upon in their pursuit of sustainable practices. However, it is important to note that this is not an exhaustive list of principles, but a list of examples of how to promote sustainability while dealing with physicalizations.

\subsubsection{Exploring sustainability from day one}
\label{subsubsec:sustainability_day_one}

To ensure the creation of a sustainable physicalization, it is important to integrate sustainability aspects right from the project's beginning. Considering \textbf{materials consistent with the data} they represent is one of the first steps while conceiving a sustainable physicalization. I5 argues that, for example, \mention{if you want to honor the ocean by making a sculpture that's inspired by the ocean, maybe don't make it out of plastic}{I5}. Alternatively, I8 deliberately chose materials that pollute the environment to highlight this issue: \mention{do we have enough material and can we make a diagram out of what actually ends up breaking down and ends up as ocean pollution?}{I8}. The \textbf{intended audience} should also be taken into consideration during the exploration phase. I8 and their collaborators, for example, focused on exhibiting their work on a specialized venue and for people that are already engaged with the topic of sustainability: \mention{our piece originally was ideally this [magazine] audience, more public interested audience who might be excited to do something}{I8}. Participants also suggest that reflecting on the physicalization's \textbf{longevity} is crucial since it affects the materials that will be used. I6 explains that \mention{the materials you choose to make a sculpture out of has a lot to do with how long it's gonna last, because wood sculptures, break down, over time. [...] Whereas people choose bronze and marble because you know they'll be around forever.}{I6}. For I1, \mention{paper was an appropriate choice}{I1} for the \project[planes-color]{Data Planes} project since it was supposed to last for a day and be recycled. The participant suggests some probing questions while reflecting on a physicalization's longevity: \mention{where do we want this to go? Where do we want this to end up? How long do we need this to exist for?}{I1}. Those and other considerations should be taken as soon as the project starts since they can greatly affect the physicalization life cycle.

\subsubsection{Creating physicalizations that break down}
\label{subsubsec:physicalizations_breakdown}

The choice of materials is a central decision for a physicalization's design. Opting for \textbf{biodegradable materials} is one solution that some of our interviewees suggested to make physicalizations more sustainable. I7, for example, explains that one of their students decided to use kombucha---biodegradable culture of yeast and bacteria---instead of long-lasting materials. Their justification is that \mention{because kombucha is a material that grows and then at one point, when it dies, if you put it in water, it just dissolves and it doesn't eject waste in the environment}{I7}. There is also a forward-looking approach to physicalization design in which the designers, artists, or researchers proactively make choices so that their \textbf{objects can break down} or they can be \textbf{dismantled} and be reused in the future. I6, for example, explains their practice: \mention{my approach to making work is in looking at craft practices because there's an insane amount of knowledge about how to live more in harmony without causing damage to the landscape. Through making objects that are meant to break down.}{I6}. S1, for example, decided to reuse the material from their prototypes by adopting this technique: \mention{using all the yarn, reduce the amount/number of trimmings. Re-using yarn by taking prototypes apart if they don't work.}{S1}. Even though destroying a physicalization is an alternative to avoid waste and environmental impact, \mention{the vast majority of materials are going to persist and we have to deal with the waste caused by them at some point}{I1}. Participants mention other practices that might help mitigate the environmental impact caused by the creation of physicalizations.

\subsubsection{Creating reusable physicalizations}
\label{subsubsec:reusable_physicalizations}

Physicalizations are commonly designed for one-off presentations, such as exhibitions or studies, restricting their broader applicability. Creating \textbf{modular} or reusable physicalization components can extend their lifespan across diverse contexts and datasets. \textbf{Reusing existing materials} or making \textbf{reusable physicalizations} was pointed out by at least 11 out of 12 survey participants as an approach used to limit the environmental impact of their physicalizations. This strategy seems to be mainly used in physicalization workshops or classes, where participants create physicalizations using simple materials that can be reused or repurposed later. S2, for instance, said: \mention{I re-use many of the same materials year-over-year, share those materials with other instructors, and repurpose them for other projects}{S2}. Physicalization kits made of durable materials are also examples of versatile physicalizations: \mention{I bring back all the tokens with me, and I'm still reusing these tokens. So these tokens were made in 2014 and I'm still using them in my class.}{I7}. Even though creating reusable physicalizations seems to be promising to mitigate their environmental impact, we could not find examples of such physicalizations designed for specific applications besides teaching or workshops. 

\subsubsection{Embracing alternative materials and practices}
\label{subsubsec:embrace_alternative}

In order to create a physicalization, designers will inevitably have to choose materials and construct physical components. Therefore, it is crucial to take into account the environmental impact of those materials and practices and choose those that have the least impact. One strategy to reduce the production of waste is working with \textbf{found materials}. During a workshop similar to the \project[physical-color]{Let's Get Physical}, I7 considered \mention{asking everybody to go out and collect some artifacts that could be garbage or that could be extracted from nature}{I7}. They justified that \mention{you can produce waste with everything right? But you can produce a physicalization with small stones that you find near a river. You don't need to have 3D printed or laser-cut tiles or whatever}{I7}. I6 also mentioned that in their art practice, they have \mention{gravitated towards wood as my primary material because it sucks to cut down trees, but it's usually like using found driftwood or like smaller pieces.}{I6}. Especially when prototyping, a similar approach is \textbf{working with scrap}. Such scrap was either purchased from second-hand stores or was found lying around in their home/studio (I2, I4, I5, and I7). Practices such as \textbf{digital prototyping} may also help the environment. Those using digital fabrication tools, mention how they \mention{prototyped digitally}{S5} and did \mention{loads of simulations before doing the 3D print}{S11} so as to reduce the waste of their process. With the same intent, S12 also optimized the cutting pattern of their CNC-ed wood structure. In summary, there are alternative practices to avoid waste during physicalization creation.

\subsubsection{Making presentation possible and engaging}
\label{subsubsec:presentation_possible}

Presenting a physicalization poses challenges in transportation, installation, and documentation. It is important to develop strategies to minimize the environmental impact during these phases while also fostering engagement with the physicalization. Many physicalizations were not built in the same place where they were presented. \textbf{Using found materials} helps with transportation since carrying around physical components might be \mention{difficult and costs money and carbon, because you are flying}{I7}. Another strategy is \textbf{using easily transportable materials}, such as \mention{a piece made of yarn, [because it] was easy to pack and transport afterwards.}{S2}. Finally, designers can also \textbf{create modular physicalizations}, as explains I5: \mention{it consists of different parts that you could connect. [...] We design it in such a way that it could fit a small truck}{I5}. Additionally, designers should also reflect on strategies to increase the engagement with the physicalization, so it could reach a large audience. I5, for example, designed part of the \project[bicycle-color]{Bicycle Barometer} \textbf{considering their audience's behavior}: \mention{we notice that cyclists tend to find things to grab on while waiting at an intersection or with their foot or with their hands. And so we thought of [designing] an armrest}{I5}. I3, on the other hand, did not have enough engagement with the \project[plants-color]{Harassment Plants} at the beginning, and had to change the \textbf{engagement strategy}: \mention{Initially we wanted to just let people engage with the physicalization themselves. But, as we did not have a good engagement, we started to invite people to look at the physicalization}{I3}. Therefore, it is necessary to plan how to present a physicalization considering the challenges related to it.

\subsubsection{Being concerned about the physicalization's end of life}
\label{subsubsec:concerned_eol}

An oftentimes unacknowledged question is ``what do we do after presenting the physicalization?''. Even though there is a growing interest in approaches to tackle the end of life of physicalizations, designers need to plan this phase carefully. A simple strategy is \textbf{repurposing the physicalization} components, as I3 did with the \project[plants-color]{Harassment Plants}: \mention{I used the pots for a while to plant stuff}{I3}. When other stakeholders participate in the design process of the project, such as in the \project[planes-color]{Data Planes} or \project[badges-color]{Data Badges}, the designer needs to be concerned about what people will do with the physicalizations afterwards. As we saw in Section \ref{subsubsec:material_ownership}, sometimes designers need to pass on the ownership of physicalizations to a partner, which may cause tensions related to sustainability. Therefore, it is necessary to \textbf{negotiate in advance the end of life} of the physicalization with other stakeholders in order to ensure they will treat it sustainably.
\section{Sustainability Dimensions for Physicalization Practice}

Our findings highlight the different perceptions that data physicalization practitioners have of sustainability. We unpack \rev{and summarize} these perceptions into ten dimensions of SDP design and present them as a multi-faceted matrix. These dimensions are structured along two main themes, \textit{intent} and \textit{impact}, the first relating to more personal value systems and the latter touching on the practical and contextual settings in which the practitioner is working. These themes and dimensions were generated through a distillation of the challenges, strategies, and considerations for sustainable physicalization identified through our findings into ten distinct considerations for SDP.

\subsection{\dimtitle{intent-color}{Intent}}

This set of dimensions relates to the designer's intent of embedding sustainability considerations in their physicalization practice (as discussed in Section~\ref{subsubsec:intent}). This includes the designer's beliefs, motivations, and aspirations for their practice and  highly reflects value-sensitive design \cite{friedman_value_nodate} approaches where participant values are accounted for in a principled and comprehensive manner throughout the design process.  We define the dimensions related to intent in physicalization practice as follows:

\begin{description}
    \item[Anticipation] relates to the attention a designer pays to integrating sustainability into their strategy and choices for a physicalization prior to initiating the project. This may comprise of the designer's general awareness and values around sustainability, as well as their awareness of and planning for the ways in which their work will have an environmental impact (some strategies are described in Section \ref{subsubsec:sustainability_day_one}). As these questions and stages are preliminary and speculative in character, techniques such as design fiction \cite{Wakkary2013} or speculative design workshops \cite{Dunne2013} might be a good starting point.
    \item[Negotiation] reflects the management of tensions or contradictions associated with designing sustainably, in order to reach an agreement between stakeholders that prioritizes a minimal environmental impact while meeting project requirements. This negotiation may relate to reaching a mutual understanding between decision-makers and participants in the project on how issues relating to sustainability will be managed. It may also describe the designer's ethos that guides their decision-making when navigating the tensions between the practicalities of the project (e.g. deadlines, budget) and designing in a sustainable way. This may be one of the most challenging aspects, as described in Sections \ref{subsubsec:material_ownership} and \ref{subsubsec:aesthetics_sustainability}. 
    Accordingly, in some cases, co-design \cite{Sanders2008} might help alleviate some of these tensions.
    \item[Inspection] is the process required to evaluate the sustainability of a physicalization project at each life cycle stage. This should include the establishment of a set of goals, metrics, or qualities that define `sustainability' for the particular project on which the designer is working, as well as a series of tests or evaluation criteria for determining if these standards have been reached. Such more quantitative metrics should not be used as success measures (with the accompanied fear of greenswashing, as described in Section \ref{sec:ethics}) but as a trigger to make personal values explicit. 
    \item[Reflection] refers to the process required to continuously reexamine, re-imagine, and improve the sustainability of one's physicalization practices. Similarly to \textit{Inspection} this reflective process should involve the designer considering the sustainability goals or values that they seek to integrate into their work, and how they might approach this integration differently in future projects.
\end{description}

\subsection{\dimtitle{impact-color}{Impact}} 

Other dimensions of SDP relate to the design choices that a designer needs to consider in order to foster sustainability in their project (as discussed in Section~\ref{subsubsec:impact}). This may include the application and setting of the physicalization, the stakeholders involved, how the design is to be presented, the available resources and budget, and the topic of the data being represented. We define these dimensions related to the impact a physicalization in the environment:

\begin{description}
    \item[Materiality] encompasses the considerations regarding the sustainability of materials and how to avoid waste during a physicalization life cycle. This might include if the material is recyclable, reusable, or biodegradable, whether the material is ethically sourced, or how much waste using the material produces (see details in Sections \ref{subsubsec:physicalizations_breakdown}, \ref{subsubsec:reusable_physicalizations}, and \ref{subsubsec:embrace_alternative}). This may involve an exploration into alternative materials (e.g. \cite{wall2021, phillips2022, rivera2023}), or the use of available sustainable materials as a driving design brief rather than a limitation.

    \item[Longevity] relates to the factors influencing how long a data physicalization persists, including the lifespan of the materials used, the rate and amount of degradation caused by usage, and the overall fragility or durability of the artifact. Additionally, designers may also consider how long the selected data will be relevant, as well as the overall topic or message of the physicalization. Recent research has highlighted the difficulties of maintenance and break-down responsibility in digital visualizations after projects are completed \cite{akbaba2023}. We believe that to be the case as well in physicalization, perhaps even more so because of the material aspects of longevity, as outlined in Section \ref{subsubsec:acknowledging_longevity}. 
    \item[Versatility] refers to the availability and ability of the physicalization (or its constituent components) to be reused for other purposes (e.g. to be used within other artifacts or to represent different data). This may include the possibility for the data to be changed or updated within the physicalization, or for physicalization to be disassembled and some or more of its parts reused in another context. We mention some challenges associated with this dimension in Section \ref{subsubsec:transience_reuse}.
    \item[Consistency] asks how the materials, tools, and production methods chosen by the designer relate to the topic or data being represented. A designer might first consider the ways in which the data they are representing is related to the values of sustainability, and consequently reflect upon how consistent their design choices for their physicalization (e.g. the materials they use, the waste produced) are with these values. Reflecting back on Mankoff's separation \cite{mankoff2007environmental}, \textit{consistency} describes the need for alignment between sustainability \textit{in} design and \textit{through} design. While this work describes the need for sustainability in design, a significant aspect of data physicalization are the topics of representation and the data which aim to persuade, educate or inform \cite{DiSalvo_2010}. We discuss this in more depth in Section \ref{sec:data}.
    \item[Visibility] is the extent to which a data physicalization reaches its intended audience, either through first-hand encounters or its dissemination through documentation. A physicalization's visibility is related to its sustainability in that it is a lens through which a designer can assess the environmental impact of their work. We describe strategies to foster physicalization visibility in Section \ref{subsubsec:presentation_possible}.
    \item[Viability] is the practical considerations that a designer needs to contend with in creating a data physicalization (e.g. budget, access to materials or equipment, availability of skills and expertise). The viability of the physicalization intersects with sustainability insomuch as it acts as a constraint on the sustainable choices a designer can make. Such challenges have also been discussed in previous physicalization work meant for public exhibitions such as~\cite{kosminsky_slave_2021}. We mention challenges related to viability in Sections \ref{subsubsec:cost_sustainability} and \ref{subsubsec:coping_transportation}. . 
\end{description}

Although we present these dimensions as distinct from each other, the realities of physicalization practices mean that there is a significant amount of overlap between them. For example, when considering \textit{Longevity} in the context of one's work, the biodegrability of a material may be an important factor, which is a key consideration within the \textit{Materiality} dimension of SDP. We see these overlaps and interlinked relationships as an inherent part of SDP.

\subsection{\rev{The Sustainable Physicalization Practices (SuPPra) Matrix}}

\begin{figure*}[!ht]
    \centering
    \includegraphics[width=0.9\textwidth]{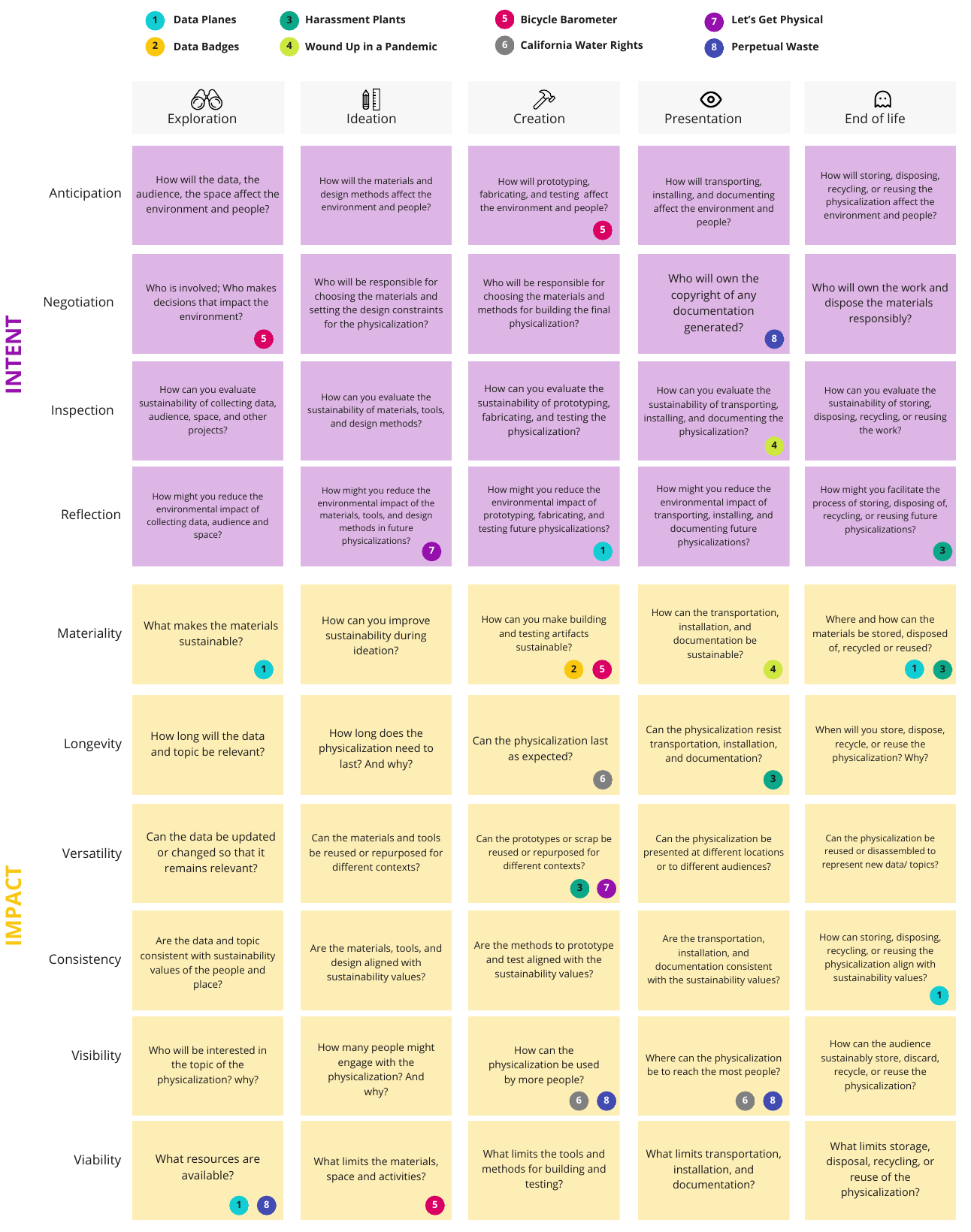}
    \centering
    \caption{\rev{The SuPPra Matrix.} The ten identified dimensions of SDP on the y-axis (as identified through our thematic analysis) mapped against the project lifecycle on the x-axis, with connections to the explored projects indicated. Note that the boundaries between these dimensions are often blurry and overlap or intersect with each other.}
    \label{fig:dimensions}
\end{figure*}

Each of the dimensions of SDP presented above offers an opportunity for physicalization designers to reflect on, re-imagine, and adapt their practices. We cross these dimensions along the different life cycle stages and create an initial scaffold for physicalization practitioners to reflect upon sustainability in their practice. In the intersections of the life cycle stages and the dimensions, as seen in Figure \ref{fig:dimensions}, we pose a set of prompting questions as a \rev{practical} tool to assist designers to plan and make choices. These questions were shaped and clarified through a reflective process in which we retroactively applied each dimension to the eight data physicalization projects associated with the interviews completed as part of this research. Accordingly, in the same Figure \ref{fig:dimensions}, we map out (using color circles) how some of these questions and considerations appeared in our project interviews. 

We imagine \rev{the SuPPra Matrix} being used as a generative lens through which a designer can view, and perhaps reconsider, their approach to sustainability, with each dimension offering a different perspective on the many decisions that make up the creation of a physicalization. We offer this matrix here as a starting point, and hope to elicit further additions by the community to create a more comprehensive overview of considerations. A high-resolution version of the matrix is available as a supplementary material.

\section{Discussion}

\rev{This section discusses the implications of our findings and outlines future directions for advancing SDP.}

\subsection{\rev{Putting Our Findings Into Perspective}}

\rev{We found that physicalization experts are aware of strategies to reduce environmental impact. Some of these practices draw inspiration from fields like HCI, Interaction Design, and Art. The transience of data---a challenge particular to the context of Data Physicalization---could be dealt, for example, with the use of electronic decomposable physicalizations to facilitate data updates~\cite{bae2022making} and subsequent decomposition of components\cite{song2022towards,song2023vim}. This strategy of breaking physicalizations down is closely related to the concepts of unfabrication~\cite{wu2020unfabricate} and unmaking~\cite{song2021unmaking}. Similarly, utilizing biodegradable materials like foam~\cite{lazaro2022exploring} or clay~\cite{bell2022reclaym} facilitates material reuse at the end of a physicalization's life cycle. Incorporating mechanical, chemical, and biological actuators into physicalizations could yield benefits comparable to electronic counterparts while potentially being more ecologically friendly~\cite{bae2022making}. Enhancing sustainability also involves repurposing components or creating reusable physicalizations. This could be achieved by reusing physicalizations `as-is', re-making them for different purposes, or remanufacturing them~\cite{kim2011practices}. Another avenue is the creation of physicalizations using found materials, similar to art projects utilizing waste materials~\cite{ezike2016exploration,somerville2017trash} or discarded items to craft innovative wearables, musical controllers, and urban interventions~\cite{moriwaki2006lessons}. We stress the significance of adapting techniques from diverse disciplines to pinpoint the most fitting strategies for lessening the environmental impact of physicalizations.}

\subsection{\rev{Can Physicalizations Be Sustainable, After All?}}

\rev{Through our interviews and survey responses, we have found numerous challenges associated with embracing sustainable physicalization practices. Sustainable materials can be costly, the transportation of large and heavy physicalizations can increase carbon footprint, and the transience of data can contribute to produce useless physical artifacts over time. This resonates with the notion that atoms are more expensive than bits~\cite{holmquist2023bits} since physicalizations are more difficult to build and manage than visualizations. This leads to the question: \textit{can physicalizations be sustainable, after all?} Lundstrom and Pargman argue that no computing system can be considered sustainable \cite{lundstrom2017developing} because they depend on the use of non-renewable resources. Holmquist also suggests that tangible user interfaces---which are close to the concept of physicalization---are not sustainable because they are significantly more expensive to create, control, modify, maintain, mass-produce, and distribute, which aligns with our findings. However, before drawing conclusions, it is crucial to emphasize the intricate relationship between sustainability and physicalizations. During our interviews, we deliberately maintained an open-ended interpretation of the term, using it as a descriptive label rather than categorizing physicalizations into a fixed `sustainable' or `unsustainable' state. Our participants expressed a similarly uncertain understanding of what the term `sustainability' truly means, leading to messy and occasionally uncomfortable conversations. Consequently, this study encountered sustainability as a `contested concept', where the fundamental understanding appears to be consensual, yet its practical implementation incites significant debate and relies heavily on contextual factors \cite{jacobs1999sustainable}.}

\rev{Despite the lack of a clear definition, our research unveiled that sustainability is a complex concept tied to various factors throughout a physicalization's life cycle. Viewing sustainability through this lens highlights its challenging nature in design due to its diverse interpretations, blurred boundaries across contexts, varied value systems, and aesthetic preferences. The fluid and evolving nature of this problem domain implies that drawing distinctions between sustainable and unsustainable physicalizations can be challenging, over-simplified and, in fact, not particularly useful. Therefore, instead of seeking rigid rules or predetermined assessments for determining whether a physicalization is (un)sustainable, this work advocates for fostering reflection throughout the life cycle of physicalizations, aligning with Remy et al.'s perspective for sustainability frameworks in SHCI~\cite{remy2017limits}. For that purpose, this work contributes with the SuPPra Matrix, which summarizes a series of reflective questions for physicalization designers structured along the lines of our findings, rather than authoritative claims. Similar to what Remy et al.~\cite{remy2018evaluation} proposed with their model, we expect that the SuPPra Matrix can empower designers and researchers to devise their own prompting questions based on their practices and contexts.}

\subsection{Data as a Dimension of Ethics and Sustainability} \label{sec:data}

We find that one aspect of Sustainable Physicalization practice that is perhaps unique to physicalization in relation to the field of HCI is its dependence on representing data. As implied in the dimension of Consistency mentioned above, we argue that a sustainable data physicalization practice should consider using ethical data. Data in and of themselves are non-tangible products of invisible labor and human effort; the outcomes of those who gather, provide, store, and own data within an opaque data system that sometimes causes harm to the environment and people involved \cite{d2020data}. For example, data storage facilities take an immense toll on the environment by consuming and harming human and ecological resources in unprecedented ways \cite{Monserrate2022Cloud}. We see the opportunities of a more transparent and accountable data that balances and upholds the well-being of the environment and its dwellers.

Our study revealed that perceptions of sustainability often overlooked the physicalization's data component as the topic of data sustainability was only spoken of indirectly. We posit that the tendency to exclude the (un)sustainability of the data used is likely because data are inherently conceptual, and the use of materials is tangible and simpler to conceptualize. Moreover, it remains difficult to select data that are sustainably produced because sustainable data provenance is largely unknown, and data governance of digital infrastructures is still developing. We thus added to the Figure \ref{fig:dimensions} the following questions that may help in projecting and speculating more possibilities in the life cycle of data physicalization: \textit{are the data represented by the physicalization sustainable? What does it mean to use sustainable data in a physicalization? How could the physical data be more sustainable through continued maintenance and reuse?} Though these prompting questions may be difficult to answer, they mark the vital need to study sustainable and ethical data sources to develop a more holistic awareness of data in HCI and Data Visualization communities. 

\subsection{Ethical Considerations for Sustainable Physicalization Advocacy}\label{sec:ethics}

Discussing sustainability is challenging because the tensions and sensitivities of sustainable practice intersect with diverse privileges and constraints that differ across people and places. Opting for a sustainable element or practice over another is an ethical and value-based decision enmeshed in socio-cultural and geographic factors. For example, all of our study participants discussed sourcing materials and choosing them based on their contexts that speak to the practical constraints they faced, such as time, funding, and other resources. Their limitations intersected with their notions of sustainability. For instance, a decision to use a seemingly non-sustainable material such as synthetic fiber over a natural or more compostable fiber was made because of budgetary limitations. However, the longevity and reuse of synthetic fiber positioned this option as sustainable when longevity was part of the designer's goals. We must acknowledge that there is no clear answer as to what is most sustainable. Still, this argument does not validate reducing efforts to improve sustainability in design practices because it is too complex to tackle.

Moreover, at the risk of our discussion being misconstrued, we must clarify that SDP is not intended as an approach to greenwashing, i.e., the practice of making a product (or physicalization in this case) appear environmentally friendly, without meaningfully reducing its environmental impact~\cite{bbc_greewashing}. We do not mean that any attempt at sustainable design is ecological and sustainable because of the complexity of sustainable design or its disparate use and meanings. Furthermore, our discussion risks portraying SDP as an approach to absolving feelings of guilt or responsibility of (un)sustainable design practices in both data and physicalization. As part of sustainability, we advocate for developing a reverence for sustainability as an access point, a gradient, a distinct context in flux, but not as an objective measure, drawing from recent work in the field \cite{bremer2022have,hansson_decade_2021}.

\subsection{Open Questions and Future Work}

The prompting questions proposed in this study were influenced by our own (originally limited) viewpoints on sustainability, as our initial questions focused mainly on the physical environment and material use. For instance, we have not touched upon broader themes of climate justice that highlight how the mitigation and responsibilities of climate change are disproportionately experienced by the most vulnerable populations and minorities~\cite{robinson2018climate,robinson2019climate}. \rev{Furthermore, the subset of physicalization experts that participated in this work were recruited primarily from the fields of research, design, and fine art. We took this approach as we believed it would provide a useful starting point for initiating the conversation around sustainable physicalization design, as well as our own familiarity and connection to these aspects of the data physicalization space. However, we should acknowledge that data physicalization as a practice spans a more diverse range of fields, including education, accessible technologies, and science communication.} While we find that our approach can serve as a good starting point, we call on our colleagues and collaborators to critically adapt and develop our \rev{SuPPra Matrix} to deepen and broaden awareness of sustainability in data physicalization \rev{practices} across many contexts. \rev{This  work is preliminary and meant to be sensitizing not prescriptive to the point that it can help quantify the environmental impact of physicalizations. There is thus a need to further bridge the practitioner perspectives described here, with expert evaluations from sustainability professionals}. In our future work, we aim to pursue ways to explore and equitably represent the different meanings and applications of sustainable design of data physicalizations.
\section{Conclusion}

This paper sheds light on the overlooked environmental impact of data physicalizations and takes significant strides in addressing this critical knowledge gap. As we continue to witness the proliferation of data-driven physical representations in various domains, it becomes imperative that we acknowledge and address the environmental impact of these practices. Through interviews with experts and a comprehensive survey, we explored the intricate relationship between sustainability and the practices associated with data physicalization. Our research revealed a multitude of sustainability considerations that stretch across the entire \textit{physicalization life cycle}. We uncovered some of the main barriers to designing physicalizations sustainably, including a lack of resources, tensions around material choices and ownership, and the transience of data. We also unveiled the strategies that physicalization \rev{experts} employ to handle these challenges, including early integration of sustainability into the project, and selecting materials based on their lifespan or reusability. Finally, we introduced ten dimensions of sustainability for physicalization practice aiming to inspire and empower physicalization designers to reflect upon sustainability in their projects. These contributions are intended to provide valuable support for researchers and practitioners as they navigate the intricacies of sustainable physicalization design.

Fostering a commitment to sustainability and responsible design principles in Data Physicalization is not merely a moral imperative but also a practical necessity. This is an initial step in considering sustainability within Data Physicalization. We hope that practitioners and researchers will draw upon the contributions outlined here to critically reflect on their future projects and explore avenues for creating more environmentally sustainable data physicalizations. We envision the community actively contributing by crafting and sharing new questions tailored to their specific contexts, thereby enriching our collective understanding of how best to address sustainability in Data Physicalization. Lastly, we look forward to witnessing the continued expansion of the field of Sustainable Data Physicalization, with further works dedicated to utilizing physicalization as a vehicle for environmental awareness and advocating for sustainability throughout the entire spectrum of physicalization practices.

\balance

\begin{acks}
We are grateful to the organizers and support of the Dagstuhl Seminar 22261, "Visualization Empowerment," where this work began. We thank our encouraging and inspiring participants and funders. This work was also supported in part by the following: Canada Research Chair in Data Visualization [CRC-2019-00368], Natural Sciences and Engineering Research Council of Canada (NSERC) Discovery Grant [RGPIN-2019-07192], and New Frontiers in Research Fund Special Call [NFRFR-2022-00570].
\end{acks}
\bibliographystyle{ACM-Reference-Format}
\bibliography{bibliography}

\end{document}